\documentclass[a4paper,10pt]{sigwebnewsletter}

\title{Manipulation and abuse on social media}

\author{Emilio Ferrara\\School of Informatics and Computing, and\\
Indiana University Network Science Institute\\Indiana University, Bloomington, IN, USA}

\newsletterQuarter{Spring}
\newsletterYear{2015}

\begin{abstract}
The computer science research community has became increasingly interested in the study of social media due to their pervasiveness in the everyday life of millions of individuals. Methodological questions and technical challenges abound as more and more data from social platforms become available for analysis. This data deluge not only yields the unprecedented opportunity to unravel questions about online individuals' behavior at scale, but also allows to explore the potential perils that the massive adoption of social media brings to our society. These communication channels provide plenty of incentives (both economical and social) and opportunities for abuse. As social media activity became increasingly intertwined with the events in the offline world, individuals and organizations have found ways to exploit these platforms to spread misinformation, to attack and smear others, or to deceive and manipulate. During crises, social media have been effectively used for emergency response, but fear-mongering actions have also triggered mass hysteria and panic. Criminal gangs and terrorist organizations like ISIS adopt social media for propaganda and recruitment. Synthetic activity and social bots have been used to coordinate orchestrated astroturf campaigns, to manipulate political elections and the stock market. The lack of effective content verification systems on many of these platforms, including Twitter and Facebook, rises concerns when younger users become exposed to cyber-bulling, harassment, or hate speech, inducing risks like depression and suicide. This article illustrates some of the recent advances facing these issues and discusses what it remains to be done, including the challenges to address in the future to make social media a more useful and accessible, safer and healthier environment for all users.
\end{abstract}

\begin{document}

\maketitle

\section*{Introduction}

Social media play a central role in the social life of millions of people every day.
They help us connect \cite{gilbert2009predicting,demeo2014facebook}, access news and share information \cite{kwak2010twitter,ferrara2013traveling}, hold conversations and discuss our topics of interest \cite{ferrara2013clustering,jafariasbagh2014clustering}.

The wide adoption of social media arguably brought several positive effects to our society: social media played a pivotal role during recent social mobilizations across the world \cite{conover2013digital,conover2013geospatial}, helping democratizing the discussion of social issues \cite{varol2014evolution}; they have been used during emergencies to coordinate disaster responses \cite{sakaki2010earthquake}; they have also proved effective in enhancing the social awareness about health issues such as obesity \cite{centola2011experimental}, or increasing voting participation during recent political elections \cite{bond201261}.

Researchers in computing quickly realized the fresh opportunities brought by the rise of social media: blending computational frameworks, machine learning techniques, and questions from social and behavioral sciences, nowadays \emph{computational social science} studies the impact of socio-technical systems on our society \cite{lazer2009life,vespignani2009predicting}.

Although a large body of literature has been devoted to the usage of social media \cite{java2007we,huberman2008social,asur2010predicting}, only recently our community started realizing some potentially harmful effects that the abuse of these platforms might cause to our society. 
Due to a mixture of social and economical incentives, a lack of effective policies against misbehavior, and insufficient technical solutions to timely detect and hinder improper use, social media have been recently characterized by widespread abuse. In the following, I illustrate what are the perils that arise from some forms of social media abuse, provide examples of the effects of such behaviors on our society, and discuss possible solutions.

\section*{Misinformation, fear, manipulation and abuse}
On Tuesday April 23rd, 2013 at 1:07 p.m. the official Twitter account of the Associated Press (AP), one of the most influential American news agencies, posted a tweet reporting two explosions at the White House allegedly causing President Barack Obama to remain injured. 
The tweet garnered several thousands retweets in a few minutes, and generated countless variants that spread uncontrolled reaching millions. 
In the short interval of time that took to other agencies to challenge the veracity of this news, and to realize that the AP Twitter account had been hacked, the panic of a terror attack started diffusing through the population; as a direct consequence, the Dow Jones plummeted 147 points in a matter of 3 minutes, one of the largest point drops in its history; shortly after the confirmation of the hack, the Dow recovered but the crash erased \$136 billion dollars. 
This event, captured in Fig. 1, is the first reported case of a planetary-scale misinformation spreading with tangible real-world effects and devastating economic loss. A later revindication by the Syrian Electronic Army raised concerns about possibly new forms of cyber-terrorism, leveraging social media, aimed at generating mass hysteria to trigger market crashes.\footnote{Syrian hackers claim AP hack that tipped stock market by \$136 billion. Is it terrorism? --- \url{washingtonpost.com/blogs/worldviews/wp/2013/04/23/syrian-hackers-claim-ap-hack}} 


\begin{figure}[!t]\centering
\label{fig:1}
\includegraphics[width=\columnwidth,clip=true,trim=4 4 4 4]{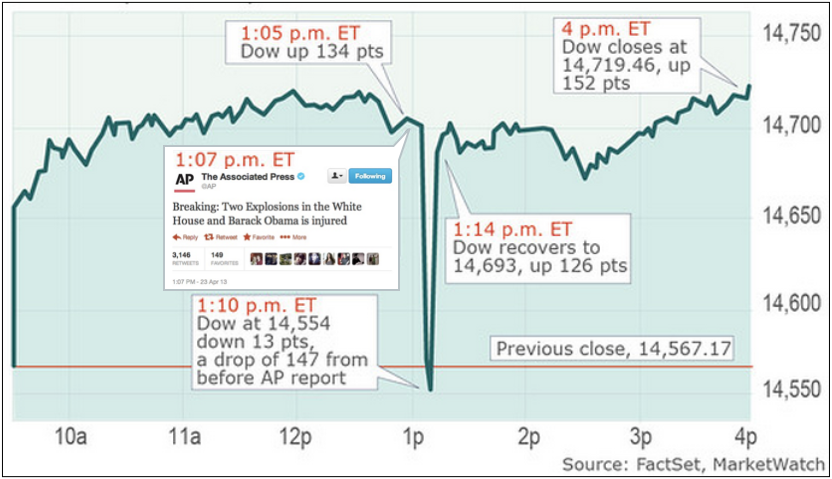}
\caption{The fake tweet causing the Dow Jones to plunge on April 23rd, 2013 in 3 minutes erased \$136 billion dollars in equity market value.}
\end{figure}

The AP hack exposed the risks related to security of social media from attacks aimed at impersonation or identity theft. 
Yet, most importantly, it showed the potentially dangerous power and effects social media conversation has on the offline, physical world ---especially on the fragile and increasingly interconnected financial markets.
This was not one isolated event: shadows have been cast \cite{hwang2012socialbots} on the role of social media during the infamous 2010 flash crash of the stock market,\footnote{Wikipedia: 2010 Flash Crash --- \url{en.wikipedia.org/wiki/2010_Flash_Crash}} after an inconclusive report by the Securities and Exchange Commission.\footnote{Findings regarding the market events on May 6, 2010 --- \url{http://www.sec.gov/news/studies/2010/marketevents-report.pdf}}
More recently, a company named Cynk Technology Corp underwent the unfathomable gain of more than 36,000\% when its penny stocks surged from less than \$0.10 to above \$20 a share in a matter of few weeks, reaching a market cap of above \$6 billions.\footnote{The Curious Case of Cynk, an Abandoned Tech Company Now Worth \$5 Billion --- \url{mashable.com/2014/07/10/cynk}} Further investigation revealed what seems to be an orchestrated attempt of touting the stock's performance through artificial social media discussion, using social bots and spam accounts \cite{ferrara2014rise,yang2014uncovering}.
The abundance of synthetically-generated content has direct implications for business ventures as well: business intelligence and analytics companies that use social media signals for market analysis (for example to predict how well a movie will do at the box office \cite{asur2010predicting,mestyan2013early}, or to determine how a new TV show is being received by it target audience) have their results affected by the noise produced by synthetic/spam accounts.\footnote{Nielsen's New Twitter TV Ratings Are a Total Scam. Here's Why. --- \url{defamer.gawker.com/nielsens-new-twitter-tv-ratings-are-a-total-scam-here-1442214842}} 

The presence of social bots on social media undermines the very roots of our information society: they can be employed to fake grassroots political support (a phenomenon called \emph{astroturf}) \cite{ratkiewicz2011truthy,ratkiewicz2011detecting}, or to reach millions of individuals by using automated algorithms tuned for optimal interaction. One recent example is provided by ISIS, which is using social bots for propaganda and recruitment \cite{berger2015isis}, adopting different manipulation strategies according to the targets of their campaigns.\footnote{The Evolution of Terrorist Propaganda: The Paris Attack and Social Media --- \url{brookings.edu/research/testimony/2015/01/27-terrorist-propaganda-social-media-berger}}
The ongoing efforts of our community to fight social bots and synthetic activity are summarized in a recent survey \cite{ferrara2014rise}.


\begin{figure}[!t]\centering
\label{fig:2}
\includegraphics[width=\columnwidth]{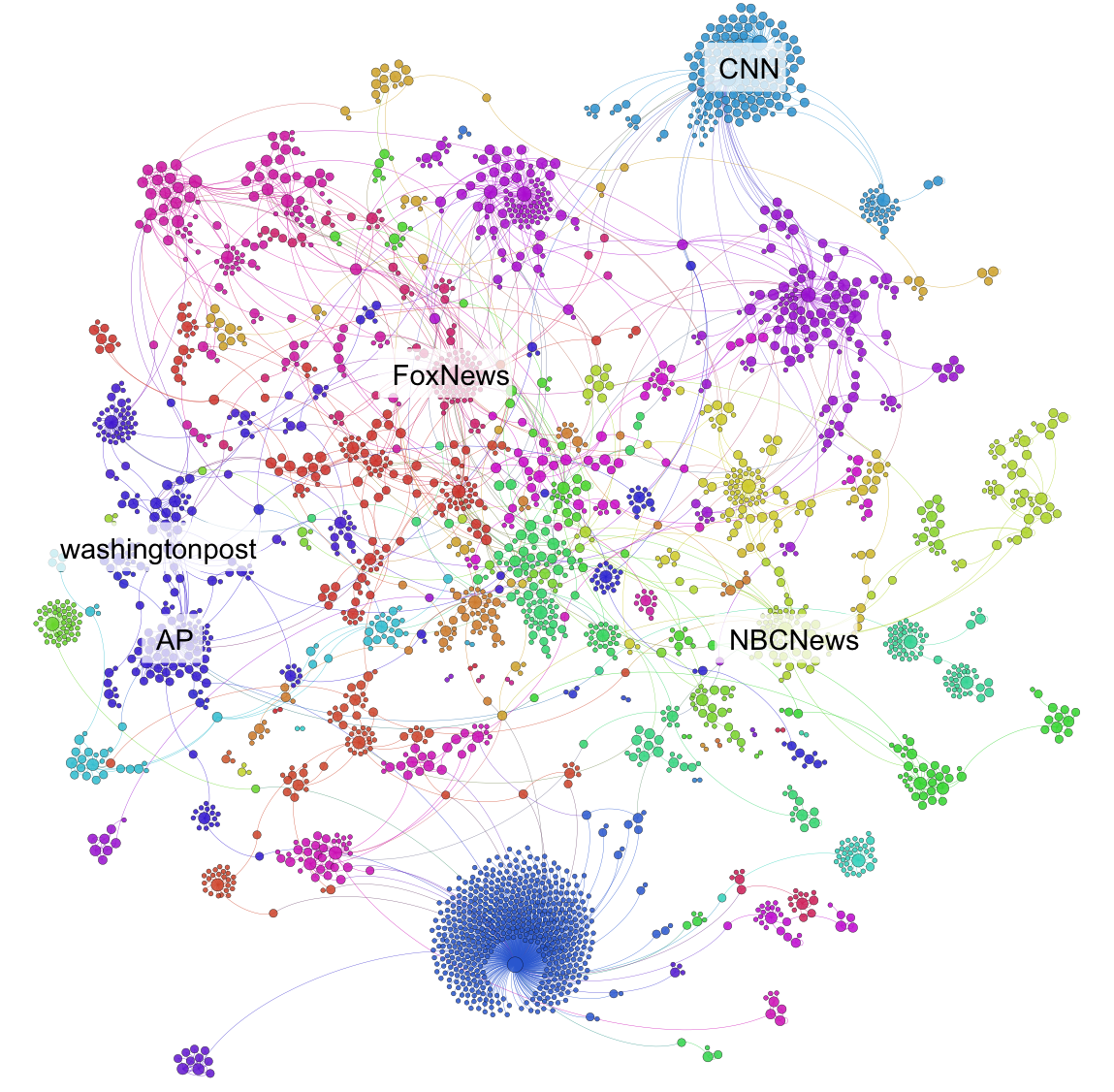}
\caption{The diffusion of information need on Twitter (RTs and MTs).}
\end{figure}

The spreading of manipulation campaigns has overwhelming societal effects. 
In politics, for example, smearing attacks have been perpetrated to defame candidates and damage their public images during various elections, including the 2009 Massachusetts Senate special election \cite{metaxas2012social} and the 2010 U.S. Senate election \cite{ratkiewicz2011detecting}, while governments and other entities attempted to manipulate the public perception on impeding social issues\footnote{Russian Twitter political protests `swamped by spam' --- \url{bbc.com/news/technology-16108876}}$^,$\footnote{Fake Twitter accounts used to promote tar sands pipeline --- \url{theguardian.com/environment/2011/aug/05/fake-twitter-tar-sands-pipeline}}.
The viral dynamics of information spreading on social media have been target of recent studies \cite{weng2013virality,weng2014predicting}, however much remains to be done to understand how to contain or hinder the diffusion of dangerous campaigns, including deception and hijacked ones.\footnote{McDonald's Twitter Campaign Goes Horribly Wrong \#McDStories --- \url{businessinsider.com/mcdonalds-twitter-campaign-goes-horribly-wrong-mcdstories-2012-1}}$^,$\footnote{NYPD's Twitter campaign backfires --- \url{usatoday.com/story/news/nation-now/2014/04/23/nypd-twitter-mynypd-new-york/8042209/}}


\begin{figure}[!t]\centering
\label{fig:3}
\includegraphics[width=\columnwidth,clip=true,trim=0 1 0 0]{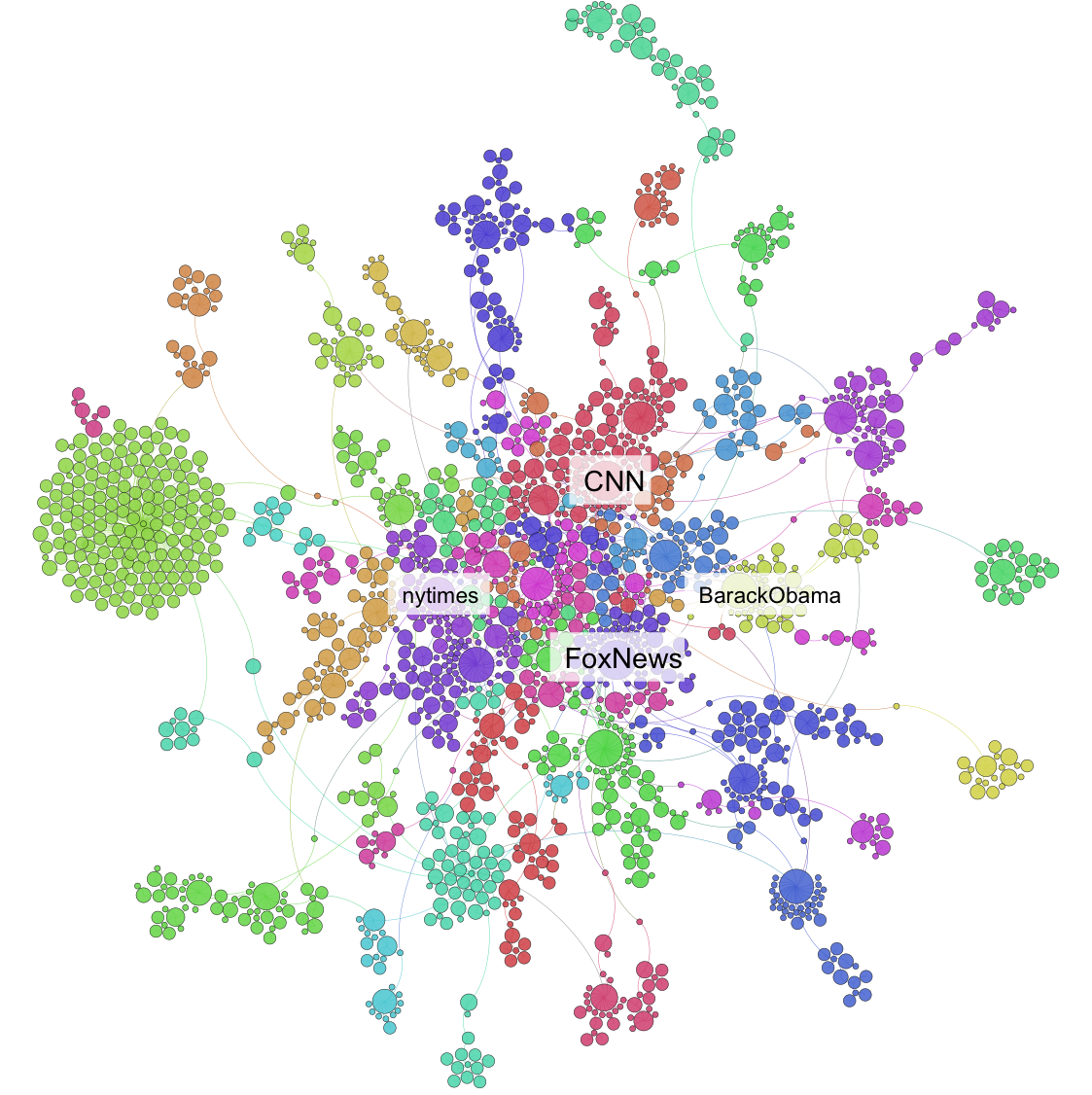}
\caption{The spreading of fear-rich content across different Twitter communities (RTs and MTs).}
\end{figure}

\begin{table}[!t]
\label{tab:1}
\centering
\begin{tabular}{ll}
\hline\hline
Time		&	Content \\
\hline \hline
2014-10-17 13:33:57 & \shortstack[l]{@FoxNews WHY WON'T YOU REPORT THE CRUISE SHIP  IS BEING \\ DENIED ENTRY INTO BELIZE? CRUCIAL FOR CLOSING OUR BORDERS!} \\
\hline
2014-10-17 13:39:44 & \shortstack[l]{@FoxNews Those hc workers are totally wacky.Y today do they decide to tell \\ some1 they handled specimens from Ebola pt, esp on ship\&not sick?} \\
\hline
2014-10-17 13:40:38 & \shortstack[l]{@NBCNews Many countries are refusing entry from flights from Ebola nations. \\ Why aren't you asking the WH about their crazy policy?} \\
\hline
2014-10-17 13:56:20 & \shortstack[l]{@CNN Why two doctors treated for EBOLA \& INFECTED no one else including \\ CARE GIVERS but it's not the same case for TX? Who screwed up?} \\
\hline
2014-10-17 13:59:20 & \shortstack[l]{@FoxNews Are they paying these potential Ebola victims money to get on ships \\ and planes? I am beginning to really wonder} \\
\hline
2014-10-17 14:09:03 & \shortstack[l]{OMG! He cld b infected "@NBCNews: Who is man wearing plain clothes \\ during an \#Ebola patient's transfer?" via @NBCDFW} \\

\hline \hline
\end{tabular}
\caption{Examples of concerned or fear-rich tweets spreading during the Ebola emergency of Oct. 17th, 2014.}
\end{table}

\bigskip

Social media have been extensively adopted during crises and emergencies \cite{hughes2009twitter}, in particular to coordinate disaster response \cite{yates2011emergency}, enhance situational awareness \cite{yin2012using}, and sense the health state of the population \cite{sakaki2010earthquake}. 
However, manipulation of information (e.g., promotion of fake news) and misinformation spreading can cause panic and fear in the population, which can in turn become mass hysteria.
The effects of such types of social media abuse have been observed during Hurricane Sandy at the end of 2012, after the Boston Marathon bombings in April 2013, and increasingly every since.
During Sandy, a storm of fake news struck the Twitter-sphere:\footnote{Hurricane Sandy brings storm of fake news and photos to New York --- \url{theguardian.com/world/us-news-blog/2012/oct/30/hurricane-sandy-storm-new-york}} examples of such misinformation spreading include rumors, misleading or altered photos,\footnote{Snopes.com: Hurricane Sandy Photographs --- \url{http://www.snopes.com/photos/natural/sandy.asp}} sharing of untrue stories, and false alarms or unsubstantiated requests for help/support.
After the Boston bombings, tweets reporting fake deaths or promoting fake donation campaigns spread uncontrolled during the first few hours after the events \cite{gupta20131}. Rumors and false claims about the capture of the individuals responsible for the bombing occurred throughout the four days after the event (the period during which the man hunt was carried out).\footnote{Wikipedia: Boston Marathon bombings ---  \url{en.wikipedia.org/wiki/Boston_Marathon_bombings}}

The examples above show how frequent social media abuse is during crises and emergencies.
My current research aims at illustrating some of the devastating societal consequences of such abuse: by extracting features of crisis-related content generated by different types of information producers (e.g., news agencies, Twitter influentials, or official organizations) along different dimensions (content sentiment, temporal patterns, network diffusion, etc.), we can study the characteristics of content produced on social media in reaction to external stimuli. Conversations can exhibit different classes of induced reactions (e.g., awareness vs. fear). We can highlight announcements and news that likely trigger positive reactions in the population (awareness), and pinpoint to those that likely yield negative feedback (fear, panic, etc.).
By identifying fear-rich content and information needs \cite{zhao2013questions}, we can characterize the relation between communication by media/organizations and emotional responses (awareness, concerns, doubts, etc.) in the population. We can study how fear and concern are triggered or mitigated by media and official communications, how this depends on language, timing, demography of the target users, etc.

Fig. 2 shows the spreading of information need \cite{zhao2013questions} on Twitter: the data capture all tweets produced during a short interval of 1 hour on October 17th, 2014, in the context of the discussion about the 2014 Ebola emergency.
Particularly prominent nodes in the discussion are labeled, and they are positioned according to their centrality; different colors identify different topical communities \cite{ferrara2013clustering,jafariasbagh2014clustering}. 
From a central core of users that start asking questions and information about Ebola, the flow of retweets (RTs) and mentions (MTs) reaches thousands, affecting even communities several hops far away from such topics, including those around influential accounts like news agencies.

Fig. 3 illustrates another example of such analysis, showing the spread of concern and fear-rich content during the same interval of time (some example tweets are reported in Table 1). 
Positioning, size and colors again represent the prominence of the accounts involved in the discussion and different Twitter topical communities. 
We can observe how panic and fear spread virally, reaching large audiences at great diffusion speed: the exposure to contents that leverage human's fears obfuscates our judgment and therefore our ability to discriminate between factual truths and exaggerated news.\footnote{Fear, Misinformation, and Social Media Complicate Ebola Fight --- \url{time.com/3479254/ebola-social-media/}}
In turn, this fosters the spreading of misinformation, rumors, and unverified news, potentially creating panic in the population, and the generation of further, more negative content. 
This self-reinforced mechanism can be interrupted by the implementation of ad-hoc intervention campaigns, designed to be effective on specific targets, based on their characteristics and susceptibility.

We urge a computational framework to deal with abuse in social media (especially during crises) that goes beyond simple data analytics, capable of actively design and implement effective policies in order to support decision-making strategies and timely interventions.

\section*{Conclusions}
In this article I illustrated some of the issues concerned with abuse on social media. Phenomena such as misinformation spreading are greatly magnified by the massive reach and pervasiveness that social media lately obtained. I brought several examples of the risks both at the economic and social level that rise from social media abuse, and discussed examples of such abuse in the context of political discussion, stock market manipulation, propaganda and recruitment. I finally exposed the consequences of the spreading of fear and panic during emergencies, as a consequence of improper communication on social media, or fear-mongering activity by specific parties. 

I highlighted some of the recent advances to challenge these issues, including the detection of social bots, campaigns, and spam, yet arguing that much remains to be done to make social media a more useful and accessible, safer and healthier environment for all users. 

Notably, recent studies illustrated how particular populations, such as younger social media users, are even more exposed to abuse, for example in the form of cyber-bullying and harassment, or being targeted by stigmas and hate speech, increasing health risks like depression and suicide \cite{boyd2008taken,boyd2014s}. 

As a community, we need to face the methodological challenges and the technical issues to detect social media abuse to secure online social platforms, and at the same time try reduce the abundant social and economical incentives by creating frameworks of effective policies against social media abuse.

\bibliographystyle{sigwebnewsletter} 
\nocite{*}
\bibliography{refs}

\begin{biography}
Emilio Ferrara is a Research Assistant Professor at the School of Informatics and Computing of Indiana University, Bloomington (USA) and a Research Scientist at the Indiana University Network Science Institute.  His research interests lie at the intersection between Network Science, Data Science, Machine Learning, and Computational Social Science. His work explores Social Networks and Social Media Analysis, Criminal Networks, and Knowledge Engineering, and it appears on leading journals like Communications of the ACM and Physical Review Letters, several ACM and IEEE Transactions and Conference Proceedings. 
\end{biography}
   
\end{document}